\documentclass[11pt,a4paper]{article}
\usepackage{amsmath, amsthm} 
\usepackage{amsfonts}
\usepackage{amssymb,graphics,psfrag}
\usepackage[bottom]{footmisc}
\usepackage{hyperref}

\def\hybrid{\topmargin -10pt    \oddsidemargin 0pt
        \headheight 0pt \headsep 0pt
       \textwidth 6.25in       
      \textheight 9.5in       
        \marginparwidth .875in
        \parskip 5pt plus 1pt   \jot = 1.5ex}
\hybrid
\numberwithin{equation}{section}
\numberwithin{table}{section}
\begin{document}

\thispagestyle{empty}

\rightline{\small MIFPA-11-33}

\vskip 3cm
\noindent
{\LARGE \bf The heterotic string at high temperature   }\\ \\
{\LARGE \bf (or with strong supersymmetry breaking)}
\vskip .8cm
\begin{center}
\linethickness{.06cm}
\line(1,0){447}
\end{center}
\vskip .8cm
\noindent
{\large \bf Waldemar Schulgin}

\vskip 0.2cm

{\em  \hskip -.05 cm George and Cynthia Mitchell Institute}
\vskip -0.15cm
{\em \hskip -.05cm for Fundamental Physics and Astronomy}
\vskip -0.15cm
{\em \hskip -.05cm Texas A \&M University, College Station}
\vskip -0.15cm
{\em \hskip -.05cm TX 77843--4242, USA}
\vskip -0.15cm
{\tt \hskip -.05cm schulgin AT physics.tamu.edu}
\vskip1cm

\noindent
{\large \bf Jan Troost}
\vskip 0.2cm
{\em \hskip -.05cm Laboratoire de Physique Th\'eorique\footnote{Unit\'e Mixte du CNRS et
    de l'Ecole Normale Sup\'erieure associ\'ee \`a l'universit\'e Pierre et
    Marie Curie 6, UMR
    8549.}}
    \vskip -.15cm
{\em \hskip -.05cm Ecole Normale Sup\'erieure}
 \vskip -.15cm
{\em \hskip -.05cm 24 rue Lhomond, 75005 Paris, France}

\vskip 1cm

\vskip0.6cm

\noindent
{\sc Abstract:} Perturbative heterotic string theory develops a single
complex tachyonic mode beyond the Hagedorn temperature.  We calculate the
quartic effective potential for this tachyonic mode at the critical
temperature. Equivalently, we determine the quartic effective
potential for strong supersymmetric breaking via anti-perdiodic
boundary conditions for fermions on a small circle.  We give many
details of the heterotic tachyon scattering amplitudes, including a
unitarity check to fix all normalization constants.  We discuss
difficulties in obtaining an effective action valid at all radii. We
argue that in certain variables, the quartic term in the potential is
radius independent.  Speculations on the properties of a new strongly
curved phase that could occur after tachyon condensation are offered.

\pagebreak

{\tableofcontents }

\newpage


\section{Introduction}

The high temperature phase of string theory is ill 
understood \cite{Atick:1988si}\cite{Bowick:1989us}\cite{Barbon:2004dd}. Beyond a critical
temperature, the canonical ensemble is no longer well-defined. There is no known
effective theory that describes the physics of strings at temperatures above the
Hagedorn temperature. The effective degrees of freedom are not known.

While the problem is universal to all perturbative string theories, we
wish to concentrate on the heterotic superstring. Due to the off-set
in the left and right level-matching conditions, the momentum and
winding of the heterotic string at finite temperature need to be
non-zero at zero oscillator excitation. Therefore, the potentially
tachyonic mode will become massive at both small and large radius.  In
fact, in the critical window only a single perturbatively tachyonic
mode can develop. That is a property unique to the heterotic string.
Moreover, the heterotic string has an enhanced gauge symmetry at
self-dual radius.

The heterotic string at finite temperature has been studied before
\cite{Atick:1988si}\cite{Alvarez:1986sj}\cite{Kounnas:1989dk}\cite{Osorio:1992vg}\cite{Dine:2003ca}. It
was argued on general grounds to exhibit a first order phase
transition. Later on, non-perturbative potentials were proposed to
describe the lower-dimensional heterotic string at finite temperature
in a spontaneous supersymmetry breaking framework
\cite{Antoniadis:1999gz} and it was shown that one can sidestep the
phase transition by introducing background potentials
\cite{Angelantonj:2006ut, Angelantonj:2008fz}.

In this paper, we will further analyze properties of the perturbative
heterotic string at finite temperature and in equilibrium.  Let us
immediately remark that this is formally equivalent to thinking of the
heterotic string as being compactified on a circle with periodic boundary
conditions for the bosonic field theory degrees of freedom, and
anti-periodic boundary conditions for the fermionic ones. We will find
the second picture more convenient for the bulk of the paper, since it
permits us to directly interpret all of the physical excitations.
Thus, we will think of heterotic string at finite temperature as a
superstring theory in which we broke supersymmetry through
 periodicity conditions (\`a la Scherk-Schwarz).

 We will calculate the quartic term in the tachyon effective potential
 at the critical radius. To that end, we plough through some of the
 detailed calculations, fix all normalizations, perform a unitarity
 check, et cetera. We give the details here since we were not able to
 extract them easily from the literature.  We then use the experience
 gained to offer speculative comments on what might happen beyond the
 Hagedorn radius.


\section{The spectrum}
In this section, we rederive the partition function of the theory under study,
and its low-energy modes in the regimes of interest to us.

\subsection{Heterotic string theory at finite temperature}
The partition function for a system at finite temperature and in
equilibrium is given by the path integral on space times a circle,
with periodic boundary
conditions for the bosonic field theoretic degrees of freedom
 and anti-periodic boundary conditions for
the fermionic ones. The inverse radius of
compactification is ($ 2 \pi$ times) the temperature (at least at
 temperatures low compared to the string scale).
Equivalently, we can think rather of a theory at zero
temperature, compactified on a circle, with twisted periodicity
conditions that will break supersymmetry. The radius of
compactification sets the supersymmetry breaking scale. When
discussing the problem at hand, we will mostly take the latter point
of view, such that we can truely pick a light-cone gauge, and talk
about all excitations as physical excitations of our system.

We determine the spectrum of the theory at generic radius of
compactification $R$.  We do this on the one hand by analyzing the
exact partition function, and on the other hand by a hands-on analysis
of the physical Hilbert space.

\subsection{The partition function}

\subsubsection{The supersymmetric ten-dimensional partition function} 

We have that the toroidal partition
function for heterotic string theory on $\mathbb{R}^{9,1}$ is:
\begin{eqnarray}
Z_{T^2} &=& i V_{10} \int \frac{d \tau d \bar{\tau}}{4 \tau_2}
\frac{1}{(4 \pi^2 \alpha' \tau_2)^5} 
\frac{1}{|\eta|^{16}}
\frac{\Gamma_{16}}{\eta^{16}} \frac{1}{2} \sum_{a,b=0}^1 
(-1)^{a+b+ab} \frac{\bar \vartheta {\left[ \begin{array}{cc} a/2  \\  b/2 
\end{array}
\right]}^4}{\bar \eta^4}.
\end{eqnarray}
There is a prefactor corresponding to the integration over $10$
zero-modes, and there are $8$ sets of
bosonic transverse oscillator excitations.  The last factor
corresponds to the GSO projected  right-moving worldsheet fermions.
The factor $(-1)^b$ corresponds to the twist by the right-moving
worldsheet fermion number while the factor $(-1)^a$ is added to
represent the minus sign due to a space-time fermion loop. The
left-movers of the heterotic superstring moreover exhibit a
sixteen-dimensional charge lattice $\Gamma_{16}$ for compact
bosons 
corresponding either to the $E_8
\times E_8$ lattice or to the $\mbox{Spin}(32)/\mathbb{Z}_2$ lattice.

\subsubsection{Compactification on a circle}

These formulas represent the partition function containing all zero
modes and all oscillator excitations transverse to the light-cone.
Compactification to a spatial circle with radius $R$ and
 periodic fields leads to
the partition function:
\begin{equation}
Z_{T^2} =2 i \pi R \, V_9 \int \frac{d \tau d \bar{\tau}}{4 \tau_2}
\frac{1}{(4 \pi^2 \alpha' \tau_2)^{5}} 
\frac{1}{|\eta|^{16}}
\frac{\Gamma_{16}}{\eta^{16}} \frac{1}{2} \sum_{a,b=0}^1 
(-1)^{a+b+ab} \frac{\bar \vartheta {\left[ \begin{array}{cc} a/2  \\  b/2 
\end{array}
\right]}^4}{\bar \eta^4} \sum_{m,w=-\infty}^{+\infty}
e^{-\frac{\pi R^2|m-w\tau|^2}{\alpha' \tau_2}}. \nonumber
\end{equation}
The quantum number $w$ corresponds to the winding and the quantum number
$m$ is Poisson dual
to the momentum. The argument of the exponential represents the classical
action of the worldsheet that  winds around the two cycles of the torus
with modular parameter $\tau$.

\subsubsection{Twisting the periodicity conditions}  
We now wish to change the periodicity conditions on the fields.
The field theory bosons remain periodic while the field theory
fermions are taken to be anti-periodic on the circle. This
breaks supersymmetry at the length scale $R$.
 It is equivalent to considering the partition
sum of the system at equilibrium and at finite temperature. In
the process, we want to retain modular invariance of the integrand
of the one loop vacuum amplitude.
These requirements lead to an extra phase factor depending
on the sector of the theory and the winding numbers of the
string \cite{Atick:1988si}:
\begin{eqnarray}
U_{ab}(m,w) &=&  \frac{1}{2}  (-1)^{ab} \Big(1+(-1)^{a+w} + (-1)^{b+m} - (-1)^{a+b+m+w}\Big).
\end{eqnarray}
It gives rise to the partition function  
\cite{Atick:1988si}\cite{Alvarez:1986sj}\cite{Kounnas:1989dk}:
\begin{eqnarray}
\!\!\!\! \!\!\!\! \!\!\!\! \!\!\!\!  Z_{T^2} &=& 2i\pi  \, R \, V_9 \int \frac{d \tau d \bar{\tau}}{4 \tau_2}
\frac{1}{(4 \pi^2 \alpha' \tau_2)^{5}} 
\frac{1}{|\eta|^{16}}
\frac{\Gamma_{16}}{\eta^{16}} \sum_{m,w=-\infty}^{+\infty}  \sum_{a,b=0}^1 
\nonumber \\
&  &\times
\frac{1}{4}
(-1)^{a+b} \Big(1+(-1)^{a+w} + (-1)^{b+m} - (-1)^{a+b+m+w}\Big)
\frac{\bar \vartheta {\left[ \begin{array}{cc} a/2  \\  b/2 
\end{array}
\right]}^4}{\bar \eta^4}
e^{-\frac{\pi R^2|m-w\tau|^2}{\alpha' \tau_2}}.
\end{eqnarray}
One can easily check that  for zero winding number $w=0$ (and for
any even $w$), there is no change due to the extra phase factor
for the bosons ($a=0$). For the fermions, one obtains an extra
factor of $(-1)^m$ in this sector, which corresponds to anti-periodic
fermions winding $m$ times. This agrees with the requirements in 
the point particle limit. The remaining phases are fixed by modular invariance.
For bosons with odd winding $w$, we find that the GSO projection
is reversed (due to an extra phase $(-1)^b$) and that an extra
phase $(-1)^m$ is added to the partition sum. After Poisson 
resummation on the summation variable $m$, we find the partition sum
in Hamiltonian form:
\begin{eqnarray} 
Z_{T^2} &=&
i V_9 \int \frac{d \tau d \bar{\tau}}{16\,  \tau_2}\,
\frac{1}{(4 \pi^2 \alpha' \tau_2)^{9/2}} \,
\frac{\Gamma_{16}}{\eta^{24} \, {\bar{\eta}}^{12}}  
\label{partitionfunction} \\
&&\times
\left(\sum_{n,w=-\infty}^{\infty} \Big(\Big(1+(-1)^w \Big)q^{\alpha' p^2_L/4}\bar q^{\alpha' p_R^2/4}+\big(1-(-1)^w\big)q^{\alpha' \tilde p^2_L/4}\bar q^{\alpha' \tilde p_R^2/4}\Big)\bar\vartheta_3^4\right.\nonumber\\
&&
-\sum_{n,w=-\infty}^{\infty} \Big(\Big(1+(-1)^w \Big)q^{\alpha' p^2_L/4}\bar q^{\alpha' p_R^2/4}+\big(-1+(-1)^w\big)q^{\alpha' \tilde p^2_L/4}\bar q^{\alpha' \tilde p_R^2/4}\Big)\bar\vartheta_4^4\nonumber\\
&&\left. -\sum_{n,w=-\infty}^{\infty} \Big(\Big(1-(-1)^w\Big) q^{\alpha' p^2_L/4}\bar q^{\alpha' p_R^2/4}+\big(1+(-1)^w\big)q^{\alpha' \tilde p^2_L/4}\bar q^{\alpha' \tilde p_R^2/4}\Big)\bar\vartheta_2^4\nonumber
\right)\nonumber
\end{eqnarray}
where the momenta are defined by:
\begin{equation}
p_{L,R} =\frac{n}{R} \pm \frac{w R}{\alpha'} \ ,
\qquad\qquad
\tilde{p}_{L,R} = \frac{n-\frac{1}{2}}{R} \pm \frac{w R}{\alpha'}\ .
\end{equation}
We see that  both the GSO projection and the momentum lattice
in the right-moving NS sector are standard at even winding number.
When the winding is odd, however, we have that the GSO projection
is reversed in the right-moving NS sector, and that the momenta
are shifted by half.

\subsection{The spectrum hands-on}

The mass shell conditions for the left-movers and
the NS and R sector right-movers are:
\begin{equation}
m^2=-\frac{4}{\alpha'}+\frac{4   N_{}^{\rm left} }{\alpha'}+\left(\frac{ n}{R}+\frac{w R}{\alpha'}\right)^2 \, ,
\end{equation}
\begin{equation}
m^2=-\frac{2}{\alpha'}+\frac{   4 N_{\rm NS}^{\rm right} }{\alpha'}+
\left(\frac{n}{R}-\frac{w R}{\alpha'}\right)^2\, ,
\end{equation}
\begin{equation}
m^2=\frac{   4 N_{\rm R}^{\rm right} }{\alpha'}+\left(\frac{n}{R}-\frac{w R}{\alpha'}\right)^2\, .
\end{equation}
The sectors refer to the sectors for the worldsheet fermion superpartners
of the ten space-time coordinates.
We will concentrate on the space-time bosons. These necessarily arise from
the right-moving NS sector. Thus we can concentrate on the left-movers,
and the NS sector right-movers. For these excitations, we necessarily have:
\begin{equation}
N^{\rm{right}}_{NS}-N^{\rm{left}}+\frac{1}{2}-nw = 0.
\end{equation}
For concreteness, we will concentrate on the $\mbox{Spin}(32) / \mathbb{Z}_2$ heterotic
superstring in the following. We introduce bosonic fields $X^A$ with
oscillators $\alpha^A$, right-moving worldsheet fermions $\psi^A$,
and 32 left-moving worldsheet fermions $\lambda^a$. The latter can be 
periodic or anti-periodic and are taken to satisfy $e^{i \pi F_\lambda}=1$,
in order to give rise to a partition function factor that is modular
invariant by itself. Note that the periodic fermions $\lambda$ do not
give rise to low-lying excitations. We will split the space-time index
$A$ into the index $\mu$ running over the non-compact directions $0,1 \dots,8$
and the index $9$, in the compactified direction.

In the left-moving sector, we have the low-lying modes:
\\
{}\\
\begin{tabular}{ccc|l}
$N^{\rm{left}}_{}$ & $m^2$ &
  & states\\ \hline&&&\\
0 & $-\frac{4}{\alpha'}+\left(\frac{n}{R}+\frac{wR}{\alpha'}\right)^2$ & 
&${|n,w\rangle^{\rm{}}_{\rm{}}}$\\[0.2cm]
1 & $\left(\frac{n}{R}+\frac{w R}{\alpha'}\right)^2$& 
&$\left( \alpha_{-1}^{\mu,9} \quad \mbox{or} \quad \lambda_{-1/2}^a\lambda_{-1/2}^b\right){|n,w\rangle^{\rm{}}_{\rm{}}}.$ \\
\end{tabular}
\vspace{.5cm}

In the right-moving NS sector we find:
\\
{}\\
\begin{tabular}{ccc|l}
$N^{\rm{right}}_{NS}$ & $m^2$ &
  & states\\ \hline&&&\\
0 & $-\frac{2}{\alpha'}+\left(\frac{n}{R}-\frac{wR}{\alpha'}\right)^2$ & 
&${|n,w\rangle^{\rm{}}_{\rm{}}}$\\[0.2cm]
$\frac{1}{2}$ & $\quad \left(\frac{n}{R}-\frac{w R}{\alpha'}\right)^2$ &
 &$\psi_{-1/2}^{\mu,9}{|n,w\rangle^{\rm{}}_{\rm{}}}.$
\end{tabular}
\vspace{.2cm}
\\ {}
\\
We tensor the left-movers with the right-moving NS sector to obtain level-matched
states in the spectrum, satisfying the modified GSO projection. We will look
for states that can be massless or tachyonic at a given radius $R$. We start
out with the left and right ground state. To level match, we need $nw=1/2$.
That gives two states labelled by (shifted) momenta
$(n,w) = \pm (1/2,1)$. They satisfy the
(reversed) GSO projection. They represent
a complex scalar $T^{\pm}$ of mass squared:
\begin{equation}
m^2=\frac{1}{4R^2}+\frac{R^2}{\alpha'^2}-\frac{3}{\alpha'}\, .
\end{equation}
The complex scalar is massless at the Hagedorn radii:
\begin{equation}
\frac{R_H}{\sqrt{\alpha'}}= 1\pm\frac{1}{\sqrt 2}\, .
\end{equation} At a radius between these
two values, the complex scalar is tachyonic, and we will loosely refer to this
scalar as the tachyon.
When we attempt to combine the right-moving ground state with the
left-moving first excited state (at level one half), we cannot satisfy
GSO and the conditions on the fermionic excitations $\lambda$.  The
same is true if we reverse left and right.  If we take both to be at
the first excited level, there will again be no solution satisfying
all consistency conditions.

The second set of states arises from the right-moving ground states, and
the left-moving state at level 1. We need $nw=-1/2$. There are two
solutions, namely $(n,w) = \pm (1/2,-1)$. The mass squared of the states is:
\begin{equation}
m^2=\frac{(-2R^2+\alpha)^2}{4R^2\alpha'^2}.
\end{equation}
The states becomes massless at the self-dual radius:
\begin{eqnarray}
\frac{R_{sd
}}{\sqrt{\alpha'}}=\frac{1}{\sqrt 2}.
\end{eqnarray}
At this radius they have left-moving momentum equal to zero
and right-moving momentum $\tilde{p}_R = \sqrt{2/\alpha'}$.
 The states represent two vectors, a complex scalar and a
complex scalar in the adjoint of $\mbox{Spin}(32)/\mathbb{Z}_2$:
\begin{equation}\left(\alpha_{-1}^\mu \quad \mbox{or}
\quad \alpha^9_{-1} \quad \mbox{or}
\quad \lambda_{-1/2}^a\lambda_{-1/2}^b\right)  \otimes | \pm\frac{1}{2},\mp1\rangle.
\end{equation}
The last possibility is to have the level $1/2$ state on the right, and the level
$1$ state on the left. We have then the constraint $nw=0$, 
and we can concentrate on the solution $(n,w)=0$. This solution corresponds to the states:
\begin{equation}
\left(\alpha_{-1}^\mu\quad \mbox{or}
\quad \alpha^9_{-1} \quad \mbox{or}
\quad   \lambda_{-1/2}^a\lambda_{-1/2}^b\right)  \otimes \left(\psi^\nu_{-1/2} \quad \mbox{or}
\quad \psi^9_{-1/2}\right) |0,0\rangle
\end{equation}
which are the $9$-dimensional graviton, dilaton and anti-symmetric
two-tensor, as well as two $U(1)$ Kaluza-Klein vectors, a scalar
radion, an scalar adjoint of $\mbox{Spin}(32)/\mathbb{Z}_2$ as well as
a vector in the adjoint of $\mbox{Spin}(32)/\mathbb{Z}_2$.  There can
be no second tachyonic mode at any radius
\cite{Antoniadis:1991kh}. Going higher in oscillation number will
automatically raise the mass squared to a positive number.

It is straightforward to check  that our hands-on description
agrees with the first terms in the expansion of the partition function
in equation (\ref{partitionfunction}).

\subsection{Summary of the low-energy excitations}

We summarize the low-energy excitations that we will be interested in.

\subsubsection{At generic radius}

We broke supersymmetry in the $\mbox{Spin}(32)/\mathbb{Z}_2$ heterotic
superstring via compactification on a circle of radius $R$ with
twisted periodicity conditions. The nine-dimensional field content of
the theory includes the nine-dimensional metric $g_{\mu \nu}$, the
anti-symmetric tensor field $b_{\mu \nu}$, the Kaluza-Klein gauge
fields $A^{g,b}_\mu$ as well as the ten-dimensional dilaton $\Phi$ and
a radion field $e^{2 \sigma} = g_{99}$. We also have the
$\mbox{Spin}(32)/\mathbb{Z}_2$ adjoint gauge field and adjoint scalar.

We also have a complex tachyon $T^\pm$ whose mass depends on the
radius of compactification. At the Hagedorn radii, it becomes massless.
In the interval between Hagedorn radii, the tachyon has negative mass
squared.

\subsubsection{At the self-dual radius}
At the self-dual radius, we find two extra massless vector fields, two
extra massless scalars as well as two extra massless $\mbox{Spin}(32)/\mathbb{Z}_2$ adjoint
scalars. The gauge group at the radius $R_{sd}=\sqrt{\alpha'/2}$ enhances
to $  SU(2)_R \times  U(1)_L\times \mbox{Spin}(32)/\mathbb{Z}_2$.  At this radius, the tachyon
left-moving momentum is $\sqrt{2/\alpha'}$ while the right-moving
momentum is zero. The tachyon is a $U(1)_L$ charged field and a
$SU(2)_R$ singlet. The two extra massless scalars that we obtain at
this radius fill out a $(3,0,1)$ multiplet \footnote{We denote the
  dimension of the group, except for $U(1)_L$, where we indicate the
  charge.} of the group $SU(2)_R \times U(1)_L  \times \mbox{Spin}(32)/\mathbb{Z}_2$. Indeed,
these states have the same transformation properties as the vectors
that fill out an adjoint of $SU(2)_R$.  The $\mbox{Spin}(32)/\mathbb{Z}_2$ adjoint scalars
combine into a $(3,0,Adj)$ multiplet.

\subsubsection{Remark}
One of our motivations to study the Hagedorn problem in the heterotic
superstring is the fact that at most one tachyon occurs at generic
radius \cite{Antoniadis:1991kh}.

\subsection{A note on spontaneous supersymmetry breaking}
While the Scherk-Schwarz reduction that we performed can be
described as a theory in which supersymmetry is broken spontaneously,
a word of caution may be needed. Indeed, the theory is supersymmetric
at infinite radius (which we can think of as the vacuum expectation
value of a scalar field) and therefore supersymmetry is broken
spontaneously in the theory, by giving a finite vacuum expectation
value to the radius field. But we must note as well that certain
fields are infinitely massive in the supersymmetric theory. Those
fields include all winding modes, and in particular the tachyon. 


From the perspective of the original ten dimensional theory these
modes are non-local. These degrees of freedom are hard to describe in
terms of a local supersymmetric Lagrangian with supersymmetry
spontaneously broken.  This phenomenon is typical of string theory (in
contrast to field theory).


\section{Scattering amplitudes} \label{scattering}

Our next goal will be to understand aspects of the effective action
for the low-energy excitations of the theory. Interesting phenomena
are bound to happen near the Hagedorn radii, where the tachyon becomes
massless, and at the  point of enhanced symmetry.
We want to reconstruct
the effective action for the low-energy excitations, and in particular
for the tachyons.
When we concentrate on the Hagedorn radius $R_H/\sqrt{\alpha'} = 1 + 1/\sqrt{2}$,
the tachyons are massless.
 The tachyons are charged under the gauge fields
arising from the reduction of the metric and the two-form. They are
moreover coupled to the graviton, the dilaton and the radion.  We will compute
the three-point amplitudes between the tachyons and the other massless
fields, to fix all normalizations in the effective action at the Hagedorn
point. We 
then also calculate the four point tachyon amplitude to determine the
coefficient of the quartic term in the effective tachyon potential.
Additionally, we will fix the overall normalization of the amplitudes
through a unitarity check. We perform the scattering amplitude
calculations for a generic radius, and restrict to the
Hagedorn point when the time has come. 

\subsection{Vertex-operators}

To compute the scattering amplitudes of tachyons and gravitational 
modes, we need their vertex operators.
The states and the 
vertex operators for the tachyons in the minus one picture are:
\begin{eqnarray}
 |0;k;n=\frac{1}{2},w=1\rangle_{-1} \  & : & \ V_{T+}^{-1}=
\frac{1}{\sqrt{2 \pi R}} e^{ik_\mu X^\mu(z,\bar z)} e^{ik_{9,L}X^9(z)+ik_{9,R}X^9(\bar z)} e^{-\tilde{\phi}(\bar z)}\ ,\nonumber\\
\!\!\!\! \!\!\!\! \!\!\!\! \!\!\!\!\!\!\!\! \!\!\!\! \!\!\!\! \!\!\!\!  \!\!\!\! \!\!\!\! \!\!\!\! \!\!\!\!   \!\!\!\! \!\!\!\! \!\!\!\! \!\!\!\! |0;k;n=-\frac{1}{2},w=-1\rangle_{-1} \  & : & \ V_{T-}^{-1}=\frac{1}{\sqrt{2 \pi R}} e^{ik_\mu X^\mu(z,\bar z)} e^{-ik_{9,L}X^9(z)-ik_{9,R}X^9(\bar z)} e^{-\tilde{\phi}(\bar z)}\, . 
\end{eqnarray}
The state-operator map maps normalized states on the left to
normalized vertex operators on the right. On the right, we have added
the extra factor $\frac{1}{\sqrt{2 \pi R}}$ because in normalizing the
vertex operator, we integrate over the compact direction.  In a
scattering amplitude, one power of the ten-dimensional string coupling
constant accompanies every vertex operator. For the tachyon, we will
take it to be $g_T^{10} = \sqrt{2 \pi R} g_T$. For the
gravitational sector we will use $g_c^{10} = \sqrt{2 \pi R}
g_c$. Unitarity requires that the equality $g_c = g_T$ holds. 
We will perform this
unitarity check.

The vertex operators 
for the metric, the dilaton, the anti-symmetric tensor,
the Kaluza-Klein gauge bosons and the radion in the minus one picture
 are:
\begin{eqnarray}\label{grvertexoperators}
\!\!\!\!\! \!\!\!\!|N=1,\tilde N=\frac{1}{2};k;n=0,w=0\rangle_{-1} \  & \! \!\!\!\ : \! \!\!\!\  & i\left(\frac{2}{\alpha'}\right)^{1/2}
\frac{1}{\sqrt{2 \pi R}}  e_{\mu\nu}\partial X_L^{\mu}(z)
\tilde\psi^\nu(\bar z)e^{ik_\rho X^\rho(z,\bar z) }e^{-\tilde \phi(\bar z)}\ , \nonumber \\
&  & i\left(\frac{2}{\alpha'}\right)^{1/2} \frac{1}{\sqrt{2 \pi R}}  e_{\mu 9}\partial X_L^{\mu}(z)\tilde\psi^9(\bar z)e^{ik_\rho X^\rho(z,\bar z) }e^{-\tilde \phi(\bar z)}\ , \nonumber \\
&  & i\left(\frac{2}{\alpha'}\right)^{1/2} \frac{1}{\sqrt{2 \pi R}}  e_{9 \mu}\partial X_L^{9}(z)\tilde\psi^\mu(\bar z)e^{ik_\rho X^\rho(z,\bar z) }e^{-\tilde \phi(\bar z)}\ , \nonumber \\
&  & i\left(\frac{2}{\alpha'}\right)^{1/2} \frac{1}{\sqrt{2 \pi R}}  e_{99}\partial X_L^{9}(z)\tilde\psi^9(\bar z)e^{ik_\rho X^\rho(z,\bar z) }e^{-\tilde \phi(\bar z)}
\end{eqnarray}
with $\mu=0,\ldots, 8$. These operators need to be normalized by choosing
polarization tensors appropriately. We take
 $e_{\mu \nu} e^{\mu \nu} =1$ as well as $e_{99} e^{99}=1$. For the radion,
we can choose $e_{99}=1$. For a Kaluza-Klein vector field associated to the
metric for instance, we can pick $e_{9 \mu} = \frac{1}{\sqrt{2}}e_\mu  = e_{\mu 9}$
and $e_\mu e^\mu=1$. 
To simplify
our notation we will refer to the Kaluza-Klein origin of these fields, and
organize parts of our calculation as a ten-dimensional one. For large parts of the
calculation, we can use ten-dimensional indices $A =0,1, \dots,9$ and
write the vertex operators as:
\begin{eqnarray}\label{vertexoperators}
V^{-1}_{T+}(z,\bar z)&=& \frac{1}{\sqrt{2 \pi R}} e^{ik^{+}_{A,L}X^A(z)+ik^{+}_{A,R}X^A(\bar z)}e^{-\tilde \phi(\bar z)}\ ,\nonumber\\
V^{-1}_{T-}(z,\bar z) &=&\frac{1}{\sqrt{2 \pi R}} e^{ik^{-}_{A,L}X^A(z)+ik^{-}_{A,R}X^A(\bar z)}e^{-\tilde \phi(\bar z)} \ ,\nonumber\\
V^{-1}_g(z,\bar z)&=&i\left(\frac{2}{\alpha'}\right)^{1/2}\frac{1}{\sqrt{2 \pi R}}
 e_{AB}\partial X^A_L(z)\tilde \psi^B(\bar z)e^{ik^{g}_{A,L}X^{A}(z)+ik^{g}_{A,R}X^{A}(\bar z)}
\end{eqnarray}
with
\begin{eqnarray}
k^{+}_{A,L/R}&=&(k_\mu,k_{9,L/R})\nonumber\\
k^{-}_{A,L/R}&=&(k_\mu,-k_{9,L/R})\nonumber\\
k^{g}_{A,L/R}&=&(k_\mu,0)
\end{eqnarray}
Note that the gravitational vertex operator $V^{-1}_g(z,\bar z)$ in
equation (\ref{vertexoperators}) is a sum of the four
operators listed in equation (\ref{grvertexoperators}).
In the following we will not always write the superscripts
$+,-,g$. Which momentum we mean should be clear from the context.
We will also need the tachyon vertex operators in the zero picture:
\begin{eqnarray}
V^{0}_{T+}(z,\bar z)&=& -\left(\frac{\alpha'}{2}\right)^{1/2}\frac{1}{\sqrt{2 \pi R}} \tilde \psi (\bar z)\cdot k_R^{+}e^{ik^{+}_{A,L}X^A(z)+ik^{+}_{A,R}X^A(\bar z)}\nonumber\\
V^{0}_{T-}(z,\bar z)&=&  \left(\frac{\alpha'}{2}\right)^{1/2}\frac{1}{\sqrt{2 \pi R}} \tilde \psi (\bar z)\cdot k_R^{-}e^{ik^{-}_{A,L}X^A(z)+ik^{-}_{A,R}X^A(\bar z)}
\label{zeropicture}
\end{eqnarray}
The dot product refers to the ten-dimensional space. We note that the
vertex operator is normalized correctly since
$k_R^{\pm \, 2}=\frac{2}{\alpha'}$. The  relative sign 
in equations (\ref{zeropicture}) makes sure that
the two-point function between these operators is positive.
 We have now enumerated all vertex
operators necessary to perform the scattering amplitude calculations.

\subsection{Amplitudes}

The calculation of the scattering amplitudes proceeds in standard fashion.
In this section, we summarize the results of these calculations.

\subsubsection{Tachyon-tachyon-gravitational three-point amplitude}

The three-point amplitudes feature two operators in the minus one picture
and one operator in the zero picture. The positions of the three
vertex operators can be gauge fixed, and give rise to ghost operator
insertions. The resulting amplitude is:
\begin{eqnarray}
{\cal A}_{T^+T^- g}&=&g_c\, g_T^2\  \left\langle c\tilde c \ V^0_{T+}(z_1,\bar z_1)\ c\tilde c \ V^{-1}_{T-}(z_2,\bar z_2)\ c\tilde c\  V^{-1}_{g}(z_3,\bar z_3)\right\rangle\nonumber\\
&=& i\pi \alpha' R \ C_{S_2} \ g_c \ g_T^2\ (2\pi)^{9}\delta^9\left(\sum_{i=1}^{3}k_i^\mu\right) e_{3\ AB} \ k_{1L}^{A+}\, k_{2R}^{B-} \nonumber \\
&=& i\pi \alpha' R\ C_{S_2} \ g_c \ g_T^2\ (2\pi)^{9}\delta^9\left(\sum_{i=1}^{3}k_i^\mu\right)\nonumber \\
& & 
\times\left(e_{3\ \mu\nu} \ k^{\mu}_1k^\nu_2-e_{3\ \mu 9} \ k^{\mu}_1k^9_R+e_{3\ 9\mu}\  k^{9}_Lk^\mu_2-e_{3\ 99} \ k^{9}_Lk^9_R\right). 
\end{eqnarray}
The overall factor $C_{S_2}$ is an overall normalization factor
consisting of the normalization of the  $X^A, \psi^A$ matter and $\tilde\phi$
ghost
path integrals.  We will fix it by demanding unitarity
\cite{Weinberg:1985tv}. To derive the amplitude we used that the
external particles are on-shell as well as the fact that pure gauge
modes decouple.  Since the graviton couples universally to tachyons,
we expect the tachyon-tachyon-graviton coupling to be the same as in
bosonic string theory, and it is \cite{Polchinski:1998rq}.

\subsubsection{Tachyon four-point amplitude}

The four-point tachyon amplitude with two negatively charged and two
positively charged tachyons features two zero and two minus one picture
operators. We order them as ${\cal A}_{T^4}=
\langle T^+(k_1) T^-(k_2) T^+(k_3) T^-(k_4)\rangle$. We integrate over the position of one of them, and the
integral gives rise to a ratio of $\Gamma$-functions:
\begin{eqnarray}
{\cal A}_{T^4}& =&-i \ (2\pi)^{2}\ C_{S_2} R \ (g_T)^4 \ (2\pi)^{9} \ \delta^9\left(\sum_{i=1}^{4}k_i^\mu\right) \nonumber\\& &  \times \frac{\Gamma\left(-1-\frac{\alpha'}{4}s\right)  \Gamma\left(-1-\frac{\alpha'}{4}t+\alpha' k_{9,L}^2\right)\Gamma\left(-1-\frac{\alpha'}{4}u\right)}{\Gamma\left(1+\frac{\alpha'}{4}s\right)\Gamma\left(3+\frac{\alpha'}{4}t-\alpha' k_{9,L}^2\right)\Gamma\left(1+\frac{\alpha'}{4}u\right)}\, .
\end{eqnarray}
The amplitude has the expected pole structure:
\begin{itemize}
\item s-channel poles at $\alpha' s=0,\ 4, \ 8, \ldots$
\item t-channel poles at $\alpha' t=4\left(-1+\alpha' k_{L,9}^2\right)=4\left(1+\alpha' k_{R,9}^2\right), \ldots$
\item u-channel poles at $\alpha' u=0,\ 4, \ 8, \ldots$
\end{itemize}
The amplitude is symmetric between the $s$ and $u$-channel.  The
lowest mass states in these channels are the modes arising
from the ten-dimensional metric, dilaton and anti-symmetric
tensor. In the $t$-channel, there can be no massless exchange, due to
momentum and winding number conservation.

\subsubsection*{Unitarity and normalization}

To compute the constant $C_{S_2}$, we use the unitarity relation which
requires that a pole in the s-channel (say) is indeed given by the
corresponding particle being exchanged through cubic vertices. We should therefore
demand that:
\begin{eqnarray}
{\cal A}_{T^4}|_{s=0} &= &-\frac{i(2\pi)^{9}\pi^2 R C_{{S_2}} (g_T^9)^4 (4+\alpha' u)(8+\alpha' u)}{\alpha' s}+\ldots \nonumber \\ 
&\!\!\!\!\!\!\!\!\!\!\!\!\!\!\!\!\!\!\!\!\!\!\!\!\!\!\!\!\!\!=&\!\!\!\!\!\!\!\!\!\!\!\!\!\!\!\!\!\!\! i\int\frac{d^9k}{(2\pi)^9}\frac{{\cal A}_{T^+ T^- G}{\cal A}_{T^+ T^- G}}{s} =-\frac{i(2\pi)^{9}\pi^2 
R^2 C_{{S_2}}^2 (g_c^9)^2 (g_T^9)^4 (4+\alpha' u)(8+\alpha' u)}{4s} + \dots\nonumber
\end{eqnarray}
which fixes the constant to be \cite{Polchinski:1998rq,Polchinski:1998rr}:
\begin{equation}\label{CS2}
C_{S_2}=\frac{4}{\alpha' g_c^2\, R  }=\frac{8\pi}{\alpha' \left(g_c^{10}\right)^2}.
\end{equation}
\subsubsection{Tachyon-Tachyon-Gravitational-Gravitational four-point amplitude}
As a cross-check on our understanding of the calculation of the amplitudes,
we wish to confirm that each vertex operator should be accompanied by
one and the same coupling constant. To that end, we compute the two tachyon
two graviton four-point amplitude and find:
\begin{eqnarray}
{\cal A}_{T^+T^- gg}&=&-i (2\pi)^{11} R C_{S_2} g_c^2 g_T^2 \ \delta^9\left(\sum_{i=1}^{4}k_i^\mu\right) \ e_{3\ AB}\ e_{4\ CD}\  \times \nonumber\\
&&\times\Big\{\alpha'\left(f(u)-1\right) k_{1L}^A\Big( \left(f(t)-1\right)k_{1L}^C+f(u)k_{2L}^C\Big)\nonumber\\
&&\times\alpha'\left(f(t)-1\right) k_{2L}^A\Big(f(t)k_{1L}^C+\left(f(u)-1\right)k_{2L}^C\Big)+2\left(f(t)-1\right)\left(f(u)-1\right)\eta^{AC}\Big\}\nonumber\\
&&\times\Big\{\alpha'\left(f(t)-1\right)k_{2R}^Bk_{1R}^D+\alpha' \left(f(u)-1\right) k_{1R}^B k_{2R}^D+2\left(f(t)-1\right)\left(f(u)-1\right)\eta^{BD}\Big\}\nonumber\\
&&\times \frac{\Gamma\left(1-f(t)-f(u)\right)\Gamma\left(-1+f(t)\right)\Gamma\left(-1+f(u)\right)}{\Gamma\left(-1+f(t)+f(u)\right)\Gamma\left(2-f(t)\right)\Gamma\left(2-f(u)\right)},
\end{eqnarray}
where the function $f$ is defined by the formula:
\begin{equation}
f(x)=-\frac{\alpha'}{4}\left(x+k_{L,9}^2\right) \, .
\end{equation}
The t-channel pole structure is:
\begin{eqnarray}
{\cal A}_{T^+T^-  gg}\Big|_{t=k_{L,9}^2-4/\alpha'}&=& -\frac{4i(2\pi)^{11} \alpha' R \, C_{S_2}\, g_c^2 \, g_T^2\  e_{4 CD} \, k_{1L}^Ak_{1R}^B\,  e_{3 AB}\, k_{2L}^Ck_{2R}^D}{t-k_{L,9}^2+4/\alpha'}\nonumber\\
&&+{\rm\ terms\  analytic \ at} \ k^2=-k_{L,9}^2+4/\alpha'.
\end{eqnarray}
This is reproduced by tachyon exchange in the $t$-channel:
\begin{equation}
i\int\frac{d^9k}{(2\pi)^9}\frac{{\cal{A}}_{T^+T^- g}\, {\cal{A}}_{T^+T^- g}}{t-k_{L,9}^2+4/\alpha'}=-\frac{i (2\pi)^{11}\alpha'^2 R^2  C^2_{S_2}\,  g_c^2\,  g_T^4 \  e_{3AB}k_{1L}^A k_{1R}^B \ e_{4CD}k_{2L}^C k_{2R}^D}{t-k_{L,9}^2+4/\alpha'}\, ,
\end{equation}
on the condition that we have $g_c=g_T$, as expected in any string theory.

\subsubsection{Summary}

The three amplitudes can be simplified, using that
 $g= g_c=g_T$ and using the value for the constant $C_{S_2}$:
\begin{eqnarray}
{\cal A}_{T^+T^- g}&=& 4i\pi  g (2\pi)^{9}\delta^9\left(\sum_{i=1}^{3}k_i^\mu\right) e_{3 AB} \ k_{1L}^{A+}k_{2R}^{B-}\label{TTG}  \, , \\
&&      \nonumber  
\end{eqnarray}
\begin{eqnarray}
{\cal A}_{T^4}&=&-\frac{16i \pi^2   g^2}{\alpha'} (2\pi)^{9} \ \delta^9\left(\sum_{i=1}^{4}k_i^\mu\right)
\nonumber \\
& & 
 \times \, \frac{\Gamma\left(-1-\frac{\alpha'}{4}s\right)\Gamma\left(-1-\frac{\alpha'}{4}t+\alpha' k_{9,L}^2\right)\Gamma\left(-1-\frac{\alpha'}{4}u\right)}{\Gamma\left(1+\frac{\alpha'}{4}s\right)\Gamma\left(3+\frac{\alpha'}{4}t-\alpha' k_{9,L}^2\right)\Gamma\left(1+\frac{\alpha'}{4}u\right)}\label{TTTT}\, ,
\end{eqnarray}
\begin{eqnarray}
{\cal A}_{T^+T^- g^2}&=&-\frac{16i \pi^2  g^2}{\alpha'}  (2\pi)^{9}\ \delta^9\left(\sum_{i=1}^{4}k_i^\mu\right) \ e_{3 AB}\ e_{4 CD}\  \nonumber\\
&&\times\Bigg(\alpha'\left(f(u)-1\right) k_{1L}^A\Big( \left(f(t)-1\right)k_{1L}^C+f(u)k_{2L}^C\Big)\nonumber\\
&&\ \ \ \ +\,\alpha'\left(f(t)-1\right) k_{2L}^A\Big(f(t)k_{1L}^C+\left(f(u)-1\right)k_{2L}^C\Big)+2\left(f(t)-1\right)\left(f(u)-1\right)\eta^{AC}\Bigg)\nonumber\\
&&\times\Bigg(\alpha'\left(f(t)-1\right)k_{2R}^B\, k_{1R}^D+\alpha' \left(f(u)-1\right) k_{1R}^B \, k_{2R}^D+2\left(f(t)-1\right)\left(f(u)-1\right)\eta^{BD}\Bigg)\nonumber\\
&&\, \times \frac{\Gamma\left(1-f(t)-f(u)\right)\Gamma\left(-1+f(t)\right)\Gamma\left(-1+f(u)\right)}{\Gamma\left(-1+f(t)+f(u)\right)\Gamma\left(2-f(t)\right)\Gamma\left(2-f(u)\right)}\, .
\end{eqnarray}
The above amplitudes are valid at any radius.


\section{The effective action}

In this section, we will use the on-shell scattering amplitudes computed
in section \ref{scattering} to fix terms in the low-energy effective action.
We start out by analyzing the effective action at the Hagedorn point(s).
We will fix cubic and quartic couplings using the scattering amplitudes.

\subsection{The action}

At the Hagedorn radius, we have the low-energy excitations arising from
Kaluza-Klein reduction from ten dimensions, supplemented with the
massless complex tachyon. We can obtain terms in the effective
Lagrangian through dimensional reduction. The reduced action
will contain a graviton, an anti-symmetric tensor, 
two scalar fields, and two Maxwell gauge fields. We then
need to couple in the tachyon. We ignore the
$\mbox{Spin}(32)/\mathbb{Z}_2$ gauge theory sector. We follow the conventions of \cite{Polchinski:1998rq}
and write the ten-dimensional string metric as:
\begin{eqnarray}
ds^2 &=& g_{\mu \nu} dx^\mu dx^\nu + g_{99} ( d x^9 + A_\mu^{g} dx^\mu)^2,
\end{eqnarray}
and we will also write the nine-nine component of the metric in terms
of a radion field $\sigma$, $g_{99} = e^{ \sigma}$. The vector field
$A_\mu^g$ is the metric Kaluza-Klein gauge field. In this section, we
take the ninth direction to be at the Hagedorn radius $x^9 \equiv x^9
+ 2 \pi R_{H}.$ 

In the Einstein frame defined by $g_{\mu\nu}^E=e^{-\frac{4\Phi_9}{D-2}}g_{\mu\nu}$
(and $D=9$),
we have the following dimensionally reduced action:
\begin{eqnarray}
S^{9\ (E)}
&=&\frac{1}{2\kappa_{9}^2}\int d^{8+1}\sqrt{-g_9^E}\Big( R^{(9)}-\frac{4}{D-2}\partial_\mu \Phi_9\partial^\mu \Phi_9-\partial_\mu \sigma \partial^\mu \sigma -\frac{1}{4}e^{2\sigma}e^{-\frac{4\Phi_9}{D-2}}F^{g}_{\mu\nu}F^{g\mu\nu}\nonumber\\
&&-\frac{1}{4}e^{-2\sigma}e^{-\frac{4\Phi_9}{D-2}}F^{b}_{\mu\nu}F^{b\mu\nu}-2 \kappa_9^2  D_\mu T^+ D^\mu T^- -2 \kappa_9^2 e^{\frac{4\Phi_9}{D-2}} V(\sigma,|T|)\Big).
\end{eqnarray}
We added in a canonically normalized
 kinetic and potential term for the complex tachyon. 
By virtue of its momentum and winding charge, the
 tachyon is charged under both Kaluza-Klein gauge fields, and the
kinetic term features gauge covariant derivatives:
\begin{equation}
D_\mu=\partial_\mu+i q^g A_\mu^{g}+i q^b A_\mu^{b} \, .
\end{equation}
We can fix part of the potential by using our knowledge of the tachyon
mass squared as a function of the radion vacuum expectation value.  We
therefore must have the following interaction term between the radion
and the tachyon:
\begin{eqnarray}
S_{m^2} &=&-\int d^{8+1}x \sqrt{-g} m^2(R = e^{\sigma} R_H) \, T^+ T^-
\nonumber \\
&=& -\int d^{8+1}x \sqrt{-g} \left(-\frac{4}{\alpha'} +
\left( \frac{1}{2R} + \frac{R}{\alpha'}\right)^2\right) T^+ T^-
\nonumber \\
&=&  -\int d^{8+1}x \sqrt{-g} \left(-\frac{4}{\alpha'} +
\left( \frac{e^{-\sigma}}{2R_H} + \frac{e^{\sigma} R_H}{\alpha'}\right)^2\right) T^+ T^-.
\end{eqnarray}
We parameterize the potential term quartic in the tachyon as follows: 
\begin{eqnarray}
S_{quartic} = -  \int d^{8+1}   \frac{\lambda}{4} |T|^4.
\end{eqnarray}

\subsection{Expansion of the action around the flat background}

We expand the action around the flat background, with constant dilaton,
the radius at the Hagedorn point, zero tachyon vacuum expectation value,
etcetera.
We will parameterize the fluctuation in the Einstein metric by:
$g_{\mu\nu}=\eta_{\mu\nu}-2\kappa_9 h_{\mu\nu}$ and will use the field
variable to parameterize the fluctuations for other fields. As a result
of expanding the effective action to cubic order, we find:
\begin{eqnarray}
S^{9\ \ {\rm cub}}&=&
\int d^{8+1}x
\Bigg(-\frac{1}{2}\partial_\mu h_{\nu\rho}\partial^\mu h^{\nu\rho} 
-2\kappa_9\Big(  h^{\mu\nu}-\frac{1}{2} {h_\rho}^\rho \eta^{\mu\nu}\Big)\partial_{\mu}T^+ \partial_\nu T^- 
\nonumber
\\
&&+\frac{1}{2\kappa_9^2}\left(-\frac{4}{D-2}\partial_\mu \Phi_9\partial^\mu \Phi_9-\partial_\mu \sigma \partial^\mu \sigma-\frac{1}{4 }F^{g}_{\mu\nu}F^{g\mu\nu}-\frac{1}{4 }F^{b}_{\mu\nu}F^{b\mu\nu}\right)\nonumber\\
&&-\frac{i}{\kappa_9^2}\Big(q^{g}A_{\mu}^{g}+q^{b}A_\mu^{b}\Big)T^{[+}\partial^\mu T^{-]}-\frac{4 \sqrt{2}}{\alpha'} 
\sigma \, T^+ T^-\Bigg) \, .
\end{eqnarray}
The normalization of the kinetic term of the gauge field and the
Kaluza-Klein momenta of the tachyon imply that we must identify the charges
in these equations as the momentum and winding around the 
Kaluza-Klein circle:
\begin{equation}\label{charges}
q^{g}=\frac{1}{2} (k_{L,9}+k_{R,9})=\frac{n}{R}\ , \qquad q^{b}=
\frac{1}{2} (k_{L,9}-k_{R,9}) = \frac{wR}{\alpha'} .
\end{equation}
In the following, we will sometimes refer to the canonically normalized
radion $\sigma'=\frac{\sigma}{\kappa_9}$ and the canonically normalized 
vector bosons  ${A}'_\mu=\frac{1}{\sqrt{2}\kappa_9} A_\mu$.

\subsection{Cubic Couplings}

We  now check the coefficients of the cubic couplings in the
effective action against the string theory three-point amplitudes
evaluated at low energy and at the Hagedorn point
$R_H/\sqrt{\alpha'}=1+1/\sqrt{2}$.  To compare the three-point
amplitudes in equation (\ref{TTG}) to the effective action, we rewrite
them using left- and right-momentum conservation, as well as the
Ward identities.
We also use the values of the ten-dimensional momenta:
\begin{eqnarray}
k_{1L}^A&=&(k_1^\mu,k_L^9)\ ,\qquad \ \ \ \ k_{2L}^A=(k_2^\mu,-k_L^9)\ ,\nonumber\\
k_{1R}^A&=&(k_1^\mu,k_R^9)\ ,\qquad \ \ \ \ k_{2R}^A=(k_2^\mu,-k_R^9)\ ,\nonumber\\
&&\qquad\quad  k_3^A=(k_3^\mu,0)\, , 
\end{eqnarray}
and we obtain
\begin{eqnarray}\label{three-pointampl}
4i \pi g_c\ e_{3\ AB}\ k_{1L}^A\ k_{2R}^B&=&2i \pi  g_c\ e_{3AB}\Big(k_{1L}^A k_{2R}^B+k_{2L}^A k_{1R}^B\Big)\nonumber\\ &\!\!\!\!\!\!\!\!\!\!\!\!\!\!\!\!\!\!\!\!\!\!\!\!\!\!\!\!\!\!\!\!\!\!\!\!\!\!\!\!\!\!\!\!\!\!\!\!\!\!\!\!\!\!\!\!\!\!\!\!\!\!\!\!\!\!\!\!\!\!\!\!\!\!\!\!\!\!\!\!\!\!\!\!\!\!\!\!\!\!\!\!\!=&\!\!\!\!\!\!\!\!\!\!\!\!\!\!\!\!\!\!\!\!\!\!\!\!\!\!\!\!\!\!\!\!\!\!\!\!\!\!\!\!\!\!\!\!\!\!\!\!\!\!2i \pi  g_c\Bigg(e_{3\mu\nu}\Big(k_1^\mu k_2^\nu+k_2^\mu k_1^\nu\Big)+{\Big(k_2^\mu-k_1^\mu\Big)\Big(k_L^9 e_{3\ 9\mu}+k_R^9e_{3\ \mu 9}\Big)}-2e_{3\ 99}k_L^9 k_R^9\Bigg)\, .  
\end{eqnarray}
To 
read the effective action terms from the amplitude, we use
the correspondence between the Fourier coefficients of the fields in
the effective action and the momentum dependent polarization tensors
used in the string amplitude computation:
\begin{eqnarray}\label{poltensor}
e_{(\mu\nu)}
-\frac{1}{D-2}{e^\rho}_\rho\delta^\perp_{\mu\nu} &\rightarrow&\  h_{\mu\nu}(k)\nonumber\\
e_{[\mu\nu ]}\ &\rightarrow&\  b_{\mu\nu}(k)\nonumber\\
\frac{1}{\sqrt{D-2}} {e^\mu}_\mu\ &\rightarrow&\  \Phi'_9(k) 
\nonumber\\
\frac{1}{\sqrt{2}}\left(e_{9\mu}\pm e_{\mu 9}\right)\ &\rightarrow&\   A_{\mu}^{'\,g/b}(k)\nonumber\\
e_{99} \ &\rightarrow&\  \sigma'(k)
\end{eqnarray}
with $\delta^\perp_{\mu\nu} =\eta_{\mu\nu}-k_{\mu} n_{\nu} - n_{\mu}
k_{\nu} $ and where a reference momentum $n$ is chosen such that it satisfies $
n^2=0$ and $k\cdot n=1$.  The normalized vertex operators  correspond
to fields $\Phi'_9$, $A_{\mu}^{'\,g/b}$ and $\sigma'$ that have canonically
normalized kinetic terms.

Next, we note that we have the
following matches between the cubic couplings and terms in the effective
action:
\begin{align}\label{amplVSaction}
4\pi i g \ e_{3\mu\nu}k_1^\mu k_2^\nu
\qquad & \rightarrow\qquad  
&-2\kappa_9\  h^{\mu\nu} \partial_\mu T^+\partial_\nu T^-\nonumber
\\\
\pi i g \ (k_{L,9}\pm k_{R,9})\left(e_{9\mu}\pm e_{\mu 9}\right) (k_2^\mu-k_1^\mu)\qquad&\rightarrow\qquad  &-2\sqrt{2}i \kappa_9 \  q^{g/b}A^{'\, g/b  \, \mu} T^{[+}\partial_\mu T^{-]}\nonumber\\
-4\pi i g \ k_{L,9} k_{R,9}\qquad & \rightarrow\qquad  &-2\kappa_9\   k_{L,9} k_{R,9}\, \sigma' T^+ T^-
\end{align}
if we identify 
\begin{equation}
2\pi g=\kappa_9.\end{equation}
Our conventions are such that the
cubic three-point amplitudes for three particles corresponding to the
fields $A_i$ are given by:
\begin{equation}
  \langle A_1(k_1)A_2(k_3)A_3(k_3)\rangle_{cubic}=\left.\frac{i\delta^3 S}{\delta A_1(k_1)\delta A_2(k_2)\delta A_3(k_3)}\right|_{ A_i =0},
\end{equation}
which applied to the effective action
 terms gives the low-energy limit of the three-point string amplitudes.

\subsection*{Remark}

We note that the couplings $\sigma \partial_\mu
T^+ \partial^\mu T^-$ and $\phi \partial_\mu T^+ \partial^\mu T^-$ do
not contribute to the on-shell three-point amplitude at the Hagedorn
point. We can fix them through the following requirement and
reasoning. Firstly, we assumed that the tachyon kinetic term is
canonical at any radius. Thus, we chose a basis of fields such that
the term $\sigma \partial_\mu T^+ \partial^\mu T^-$ is absent from the
Lagrangian.  Otherwise, a non-zero vacuum expectation value for the
radion $\sigma$ would lead to a non-canonical kinetic term for the
tachyon.  Secondly, by the universality of the dependence of the
string effective action on the dilaton, we note that after going to
Einstein frame, the tachyon kinetic term is dilaton independent. Thus,
the term $\phi \partial_\mu T^+ \partial^\mu T^-$ is also absent from
the Lagrangian.

\subsection{A Quartic Coupling}

In this subsection, we want to determine the quartic term in the tachyon potential.
The tachyon four-point amplitude
in string theory, expanded under the assumptions 
$u \alpha' <<1, s \alpha'<<1$ gives rise to poles,
constant terms and terms of higher order in the momenta:
\begin{eqnarray}\label{expansionampl}
  {\cal A}_{T^4} & \approx &
  -\frac{32i\ \kappa_9^2  }{\alpha'^2}
  \left(\frac{3}{4}\alpha' +\frac{1}{s} \left(1+ \frac{\alpha'}{4} u\right)\left(1 + \frac{\alpha'}{8} u\right)
    + \frac{1}{u} \left(1+ \frac{\alpha'}{4} s\right)\left(1 + \frac{\alpha'}{8} s\right)\right)
  +\dots
  \nonumber \\
  & \approx & -i \kappa_9^2  \left(\frac{24}{\alpha'}+
    \left(\frac{u^2}{s} + \frac{s^2}{u}\right) 
    + 
    \frac{12}{\alpha'} \left( \frac{u}{s} + \frac{s}{u}\right)
    +
    \frac{32}{ \alpha'^2} \left(\frac{1}{s}+ \frac{1}{u}\right) + \dots
  \right).
\end{eqnarray}
As a cross-check on the overall sign, we note that the graviton
exchange is as in the Virasoro-Shapiro amplitude in bosonic string
theory \cite{Polchinski:1998rq}. The pole contributions of the
amplitude correspond to the exchange of massless particles. These
exchange contributions can also give rise to constant contributions
that we need to subtract in order to isolate the contribution from the
quartic contact term. Above we have fixed a choice of fields and
associated cubic couplings, so that the exchange contribution is now
fixed. Let's compute it.

\subsubsection*{Massless exchange contributions to the four-point tachyon amplitude}

The massless states that can be exchanged by the tachyons are the
graviton, the radion, the dilaton and the two gauge bosons. We compute
the amplitude for the exchange diagrams using the expression
\begin{equation}
{\cal A}_{A_1A_2\rightarrow B_1B_2}=i \frac{\delta^3 S}{\delta A_1(k_1)\delta A_2(k_2)\delta A_3(k_3)} \ \Delta^{-1}_{A_3A_3}\ i \frac{\delta^3 S}{\delta A_3(k_4)\delta B_1(k_5)\delta B_2(k_6)}
\end{equation}
for the exchange amplitude, 
where $\Delta^{-1}$ is the propagator of the exchanged state.  Due to
the fact that both the $\sigma \partial_\mu T^+\partial^\mu T^-$ and
the $\Phi_9 \partial_\mu T^+\partial^\mu T^-$ term are absent in our
effective Lagrangian, we can
concentrate on the non-derivative cubic terms in the action for the radion
and the dilaton. Due to
the fact that the tachyon is massless, there is moreover no $\Phi_9
T^+ T^-$ term in the action. We therefore can concentrate on the
exchange of the graviton, the radion and the two gauge fields.
\begin{itemize}
\item We compute the graviton exchange in the harmonic gauge using the propagator:
\begin{equation}
\Delta^{-1}_{\mu\nu,\rho\sigma}=-i\Big(\frac{1}{2}\eta_{\mu\rho}\eta_{\nu\sigma}+\frac{1}{2}\eta_{\mu\sigma}\eta_{\nu\rho}-\frac{1}{D-2}\eta_{\mu\nu}\eta_{\rho\sigma}\Big)\frac{1}{k^2}.
\end{equation}
The $s$-channel contribution is:
\begin{eqnarray}
{\cal A}_{T^+T^-\rightarrow T^+T^-}^{\rm {s-channel}}&=& \Bigg(2i \kappa_9 \Big(k_1^\mu k_2^\nu-\frac{1}2{}\eta^{\mu\nu}k_1\cdot k_2\Big)\Bigg)\ \Delta^{-1}_{\mu\nu,\rho\sigma} \ \Bigg(2i \kappa_9\Big( k_3^\rho k_4^\sigma-\frac{1}{2}\eta^{\rho\sigma}k_3\cdot k_4\Big)\Bigg)\nonumber\\
&=& 
-i \kappa_9^2\left(\frac{u^2}{s}+u\right).
\end{eqnarray}
Summing over $s-$ and $u$-channels gives:
\begin{equation}
{\cal A}_{T^+T^-\rightarrow T^+T^-}^{\rm {graviton\ exchange}}=-i \kappa_9^2\left(\frac{u^2}{s}+\frac{s^2}{u}+s+u\right).
\end{equation}
\item The terms in the action relevant to the normalized radion exchange
are:
\begin{eqnarray}
S
=-\frac{1}{2} \int d^{8+1} x \Bigg( \partial_\mu \sigma' \partial^\mu \sigma'+
\kappa_9 \frac{8 \sqrt{2}}{\alpha'} \sigma' T^+ T^-\Bigg).
\end{eqnarray}
The radion propagator is $\Delta_{\sigma'}=-\frac{i}{k^2}$, such that we obtain
the radion exchange contribution to the tachyon four-point amplitude:
\begin{eqnarray}
{\cal A}_{T^+T^-\rightarrow T^+T^-}^{\rm {radion\ exchange}} &=& 
- \frac{32 i\  \kappa^2_9}{\alpha'^2} \left (\frac{1}{s} + \frac{1}{u}\right).
\end{eqnarray}
\item Both the metric gauge field $A^{g}$ and the anti-symmetric
  tensor gauge field $A^{b}$ are exchanged. Since the terms in the
  action are the same up to the charge, we find an
  exchange contribution proportional to  the
  respective charges squared.
The canonically normalized fields have an action equal to:
\begin{equation}
\int d^{8+1}x\Bigg(- \frac{1}{4}F_{\mu\nu}^{'\ g}F^{'\ g\mu\nu}-
\sqrt{2} i \kappa_9  \ q^{g}A_\mu^{'g}\Big(T^+\partial^\mu T^--T^-\partial^\mu T^+\Big)\Bigg).
\end{equation}
Using the propagator
\begin{equation}
\Delta^{-1}_{\mu\nu}=-\frac{i\eta_{\mu\nu}}{k^2},
\end{equation}
we then find the total contribution of both gauge bosons to the
tachyon four-point amplitude summed over $s$- and $u$-channel to be:
\begin{equation} {\cal A}_{T^+T^-\rightarrow T^+T^-}^{\rm {boson\
      exchange}} = - 2i \kappa_9^2 \left(q_g^2+q_b^2\right)
  \left(\frac{2s}{u}+\frac{2u}{s}+ 2\right)=-  i
  \kappa_9^2 \ \frac{12}{\alpha'}\left(\frac{s}{u}+\frac{u}{s}+ 1\right)\ ,
\end{equation}
where in the last step we inserted the value of the charges at the 
Hagedorn radius $q^{g/b}=\frac{1}{\sqrt{\alpha'}}\left(1\pm \frac{1}{\sqrt{2}}\right)$.
\item In summary, the sum of the exchange contributions of all
  massless states in both channels are:
\begin{equation}\label{exchangediagrams}
{\cal A}_{T^+T^-\rightarrow T^+T^-}^{\rm {all\ exchanges}} =
-i \kappa_9^2\Bigg(\left(\frac{u^2}{s}+\frac{s^2}{u}+s+u\right)+\frac{32}{\alpha'^2} \left (\frac{1}{s} + \frac{1}{u}\right)+ \frac{12}{\alpha'} \left(\frac{s}{u}+\frac{u}{s}+ 1\right)\Bigg).
\end{equation}
\end{itemize}
The contributions match the terms in the expansion of the amplitude
$A_{T^4}$ in equation (\ref{expansionampl}) near small momenta $u \alpha' <<1, s
\alpha'<<1$.  Note that the gauge boson exchange gives rise to a
constant term. Indeed, the pole can be cancelled by the two cubic
one-derivative interactions. We also remark that the scalars don't
give rise to a constant term because of the absence of cubic
one-derivative couplings.

\subsubsection*{The quartic term in the tachyon potential}

The contact contribution to the amplitude
 arising from the quartic term in the effective action
for the tachyon
is given by:
\begin{eqnarray}
-i \lambda.
\end{eqnarray}
To obtain the value of the coupling constant $\lambda$ we subtract
the contributions from the exchange diagrams in equation (\ref{exchangediagrams})
from the expansion of the amplitude $A_{T^4}$ in equation
(\ref{expansionampl}) near small momenta. We find:
\begin{equation}
\lambda= \frac{12\, \kappa_9^2 }{\alpha'}.
\end{equation}
The potential due to the contact term is a repulsive delta-function
potential.

\subsubsection*{Remark}

We believe it is an interesting result to have determined the quartic
tachyon potential in our choice of field basis. For the order of the
phase transition at the Hagedorn point though, the cubic interaction
between the radion and the tachyon is sufficient to argue it is of
first order \cite{Atick:1988si}.  One reason to nevertheless go
through this elaborate determination of the quartic potential at the
Hagedorn point is that it can be determined rigorously at this point.
We will attempt to draw lessons from this exercise in the next, more
exploratory section.


\section{The effective action at smaller radii}

In this section, we want to speculate about what happens at smaller
radii, and in particular at the most symmetric point, namely the
self-dual radius. We face a generic problem that plagues our
understanding of the physics at the self-dual radius, which is that
on-shell, the tachyon momentum will be of order one over the string
scale, thus rendering invalid the derivative expansion in the
effective action. Only for very short time scales would one be able to
trust the effective action near flat space.  In this section, we
therefore concentrate on enumerating some generic expected properties
of the effective action, and we will explore one avenue to get a
handle on the physics at small radii.

\subsection{The effective action at the self-dual radius}

At self-dual radius, we will have an enhanced $SU(2)_R \times U(1)_L
\times \mbox{Spin}(32)/\mathbb{Z}_2$ gauge symmetry.  The radion joins
a $(3,0,1)$ triplet, and the $\mbox{Spin}(32)/\mathbb{Z}_2$ adjoint
scalar also becomes part of a triplet. The $SU(2)_R$ triplet will
presumably give rise to a quartic commutator square potential. The
triplet and the tachyon will also give rise to a quartic term, and we
may expect a potential in terms of the tachyon only as well. We can
also have a commutator squared term for a mix of the two $SU(2)_R$
triplets. We note that the cubic potential term $ \sigma T^+ T^-$ is
absent at the self-dual radius, since the self-dual radius corresponds
to the minimal value of the tachyon mass squared.  

For concreteness, let us focus on the term in the potential that
depends on the complex tachyon only.  For starters, we note that the
tachyon vertex operator in the zero picture has a term proportional to
the space-time momentum, and a term proportional to the chiral compact
momentum $p_R$ on the supersymmetric side of the heterotic string. The
momentum $p_R$ is zero at the self-dual point. Furthermore, the
tachyon $N$-point function with $N$ larger or equal than two contains
at least one zero picture tachyon vertex operator. Therefore, all the
$N$-point amplitudes will be proportional to at least one factor of
the space-time momentum. In fact, by Lorentz symmetry, they will be
proportional to at least the space-time momentum squared.  However, we
cannot use this property to exclude a tachyon potential, since the
integration over the positions of the tachyon vertex operators leads
to extra poles in the amplitude. We will proceed differently to understand
how the tachyon potential depends on the radius.

\subsection{A universal behaviour}

In fact, if we attempt to identify universal properties of the tachyon
four-point amplitude at any radius, we can proceed as follows. Since the
mass squared of the tachyon is radius dependent, we are tempted to
eliminate it using the on-shell relation\footnote{Our procedure is inspired by
\cite{Garousi:2003pv}.}. After this elimination, we wind up
with a four-point amplitude: 
\begin{equation}
 {\cal A}_{T^4} = -\frac{16i \pi^2   g^2}{\alpha'} (2\pi)^{9}\ 
\delta^9 \left(\sum_{i=1}^{4}k_i^\mu \right)  \frac{\Gamma\left(-1-\frac{\alpha'}{4}s\right)
\Gamma\left(1+\frac{\alpha'}{4}(s+u) \right)
\Gamma\left(-1-\frac{\alpha'}{4}u\right)}{
\Gamma\left(1+\frac{\alpha'}{4}s\right)
\Gamma\left(1-\frac{\alpha'}{4}(u+s) \right)
\Gamma\left(1+\frac{\alpha'}{4}u\right)} \label{TTTTuni}
\end{equation}
which depends on the Mandelstam invariants $s$ and $u$ only.
These are the Mandelstam variables which exhibit massless poles.
 The only
radial dependence in the amplitude is an overall factor. Since the
amplitude no longer depends on the mass squared of the tachyon, we can
attempt to interpret this amplitude as being off-shell. We could think
of it as a function of $s,t,u$ which happens to be independent of
$t$, with no kinematic constraint on the variables. A justification of
this off-shell extension would have to come from a heterotic string field theory.

We now wish to venture even further, and we will attempt to identify the quartic
tachyon potential term coded in this amplitude.

\subsection{Field theory exchange diagrams}

We will take a pedestrian approach here, and discuss the same field
theory exchange diagrams that we already identified before, in an
optimistic analogy to the effective action approach valid at the
Hagedorn radius. We do take into account the change in the mass
squared of the tachyons, the extra couplings this generates to the
dilaton, the trace of the metric, and the change in the gauge
couplings of the tachyon. We also note that the tachyon is an $SU(2)_R$
singlet, which dictates that no new massless gauge bosons are exchanged.

The cubic tachyon coupling to the graviton, and to the dilaton
acquire extra terms that can be read off from the expansion:
\begin{eqnarray}
\!\!\!\!\!\!\!\!\!\!\!\!\!\!\!S^{9(E)}&=&\int d^{8+1}x\sqrt{-g}\left(-\frac{1}{\kappa_9^2} \frac{2}{D-2}\partial_\mu \Phi_9 \partial^\mu \Phi_9-e^{\frac{4\Phi_9}{D-2}}m^2 T^+T^-+\ldots \right)\nonumber\\
&\approx&
\int d^{8+1}x \Bigg(-\frac{1}{2}\partial_\mu \Phi'_9\partial^\mu \Phi'_9-\frac{2m^2}{\sqrt{D-2}}\kappa_9 \Phi'_9 T^+T^-+\kappa_9 m^2 {h^\mu}_\mu T^+ T^- +\ldots  \Bigg)\, ,
\end{eqnarray}
with $\Phi'_9=\frac{1}{\kappa_9} \frac{2}{\sqrt{D-2}}\Phi_9$ (and $D=9$ in our case).
The cubic coupling to the dilaton and the trace of the metric are proportional
as determined by the universal prefactor $e^{- 2 \Phi_9} \sqrt{-g}$ to the potential
term in the string frame. 
The full cubic vertices become:
\begin{eqnarray}
V_{T^+T^-h}&=&2i\kappa_9 \left( 
k_1^\mu k_2^\nu-\frac{1}{2}\eta^{\mu\nu}(k_1\cdot k_2-m^2) \right) \, ,\nonumber\\
V_{T^+T^-\Phi'_9}&=&-\frac{2im^2 \kappa_9}{\sqrt{D-2}}\, .
\end{eqnarray}
The graviton and the dilaton exchange then sum up to:
\begin{eqnarray}
{\cal A}^{{\rm h+\Phi'\ exchange}}&=&-2i \kappa_9^2\Bigg(\frac{1}{s}\Big(k_1\cdot k_3\ k_2\cdot k_4+k_1\cdot k_4\  k_2\cdot k_3-\left(k_1\cdot k_2-m^2\right)\left(k_3\cdot k_4-m^2\right)\Big)\nonumber\\
&&
+\frac{1}{u}\Big(k_1\cdot k_3\ k_2\cdot k_4+k_1\cdot k_2\ k_3\cdot k_4-\left(k_1\cdot k_4-m^2\right)\left(k_2\cdot k_3-m^2\right)\Big)\Bigg)\nonumber\\
&=& -i\kappa_9^2\left(\frac{u^2}{s}+\frac{s^2}{u}+u+s+4m^2\left(-1-\left(\frac{u}{s}+\frac{s}{u}\right)+m^2\left(\frac{1}{s}+\frac{1}{u}\right)\right)\right).
\end{eqnarray}
Note that $D$-dependence drops out.
The gauge boson exchange gives rise to the amplitude:
\begin{eqnarray}
{\cal {A}}^{{\rm A\ exchnage}}&=&-4i\kappa_9^2\left(q_g^2+q_b^2\right)\left(\frac{1}{s}(k_1\cdot k_3-k_1\cdot k_4)+\frac{1}{u}(k_1\cdot k_3-k_1\cdot k_2)\right)\nonumber\\
&=&-i \kappa_9\left(\frac{12}{\alpha'}+4m^2\right)\left(1+\frac{u}{s}+\frac{s}{u}-2m^2\left(\frac{1}{s}+\frac{1}{u}\right)\right).
\end{eqnarray}
The radion exchange contribution becomes:
\begin{eqnarray}
{\cal A}^{{\rm \sigma\ exchnage}}&=&-4i \kappa_9^2 \ k_L^2 k_R^2\left(\frac{1}{s}+\frac{1}{u}\right)=-4i\kappa_9^2\left(m^2+\frac{4}{\alpha'}\right)\left(m^2+\frac{2}{\alpha'}\right)\left(\frac{1}{s}+\frac{1}{u}\right)\nonumber\\
&=&-i\kappa_9^2\left(4m^4+\frac{24m^2}{\alpha'}+\frac{32}{\alpha'^2}\right)\left(\frac{1}{s}+\frac{1}{u}\right).
\end{eqnarray}
All exchanges summed up give:
\begin{equation}\label{exchangediagramsM}
{\cal A}_{T^+T^-\rightarrow T^+T^-}^{\rm {all\ exchanges}} =
-i \kappa_9^2\Bigg(\left(\frac{u^2}{s}+\frac{s^2}{u}+s+u\right)+\frac{32}{\alpha'^2} \left (\frac{1}{s} + \frac{1}{u}\right)+ \frac{12}{\alpha'} \left(\frac{s}{u}+\frac{u}{s}+ 1\right)\Bigg) .
\end{equation}
We note that the mass squared dependence of the exchange diagrams
drops out in the total exchange contribution. For the singular terms
in the Mandelstam variables $s$ and $u$, this has to be the case, as
they need to reproduce the mass squared independent singular terms in
the string scattering amplitude. We find that also the constant term
is independent of the mass squared in our frame.

\subsubsection*{Universality}

As a consequence of the independence of the mass squared of both the
amplitude and the exchange contribution as a function of the
Mandelstam variables $s$ and $u$, the quartic tachyon potential
contact term will remain $\lambda=12 \kappa_9^2/\alpha'$ at all radii.
One needs more information to conclude that there is a 
new locally stable minimum (e.g. at the self-dual radius).

\section{Conclusions and speculations}

We have reviewed the spectrum of the heterotic superstring
at high temperature compared to the string scale. Equivalently,
we have analyzed the theory after compactification on a circle
with supersymmetry breaking periodicity conditions. We determined
the coefficient of the quartic term in the tachyon potential
at the Hagedorn radius. The
calculation was presented in great detail, including normalized
vertex operators, a unitarity check, and careful subtraction of the
exchange diagrams. 

We argued that a description of the physics in terms of a
spontaneously broken supersymmetric theory needs to take into account
winding modes that generically will not fit into supermultiplets. This
phenomenon is typical of string theories.  We also showed that 
naively extending the range of validity of the effective action to
string scale regions leads us to a tachyon quartic potential that has
a coefficient independent of the radius of compactification.

If we let both the radion and the dilaton fluctuate, then Yukawa type
interactions will ensure a first order phase transition.  If we fix
the radius at the self-dual value, and ignore the runaway dilaton, we
find a quartic tachyon potential with string scale symmetry breaking
minima.  Massive stringy modes, higher tachyon potential terms, and
higher derivative terms will then become important and they could
either stabilize or destabilize the new phase.

If we take the view where we compactify a space-like circle with
supersymmetry breaking periodicity conditions, then the winding
tachyon is a straightforward physical excitation that can obtain a new
vacuum expectation value. We find it interesting to speculate that
below the Hagedorn radius, the tachyon might condense entirely
homogeneously in the rest of space-time, leading to a winding-momentum
condensate of fundamental strings in the compactified direction.  The
true vacuum would have at least a $SO(8,1)$ space-time isometry. The
$U(1)_L$ of the circle would be broken by a tachyon condensate, and
the $SU(2)_R$ would be unbroken.  In the new phase, the energy of
space-time is lower than in the old phase, and we might expect a
homogeneous negative cosmological constant. A new background of the
form $AdS_9$ at string scale is consistent with these
expectations. There would be space-time symmetry enhancement. The
background would be complemented with a tachyon expectation value.
Massive string modes and Kaluza-Klein excitations would mix, and would
result in a stringy geometry.  One can attempt to demonstrate the
existence of these backgrounds directly. See e.g.
\cite{Polyakov:2005ss,Friess:2005be}.

To get a handle on this phase it might be necessary to have a
practical formulation of heterotic string field theory. Since for the
heterotic string the most tachyonic direction is easily identified by
spectral analysis, the problem seems worthy of attention.
String field theory was extremely useful in furthering our
understanding of open string tachyon condensation \cite{Sen:1999mg,Sen:1999nx}.

Another avenue of research would be to analyze in this
Scherk-Schwarz model how tachyon condensation influences ultraviolet properties of the spectrum through the UV/IR
connection in the closed string torus amplitude \cite{Kutasov:1990sv, Angelantonj:2010ic} \footnote{We would like to thank Carlo Angelantonj for a
clarifying discussion on this point.}.


\section*{Acknowledgements}

We would like to thank Carlo Angelantonj, Costas Bachas, Mohammad
Garousi, Guangyu Guo, Costas Kounnas, Daniel Robbins and Nick Toumbas
for interesting discussions and correspondence.  Our work was
supported in part by the grant ANR-09-BLAN-0157-02. The work of WS was
supported by NSF under PHY-0505757 and the University of Texas
A\&M. WS would like to thank Ecole Normale Sup\'erieure for kind
hospitality during different stages of this project.

\bibliographystyle{utphys}
\bibliography{hagedorn}

\end{document}